# STUDY OF THE MAGNETIZING RELATIONSHIP OF THE KICKERS FOR CSNS*

M.Y. Huang[1,2#], Y.W. An[1,2], S.N. Fu[1,2], N. Huang[1], W. Kang[1,2], Y.Q. Liu[1,2], L. Shen[1,2], L. Wang[1,2], S. Wang[1,2], Y.W. Wu[1,2], S.Y. Xu[1,2], J. Zhai[1,2], J. Zhang[1,2]
[1]Institute of High Energy Physics, Chinese Academy of Sciences, Beijing, China
[2]Dongguan Neutron Science Center, Dongguan, China

*Abstract*

The extraction system of CSNS mainly consists of two kinds of magnets: eight kickers and one lambertson magnet. In this paper, firstly, the magnetic test results of the eight kickers were introduced and then the filed uniformity and magnetizing relationship of the kickers were given. Secondly, during the beam commissioning in the future, in order to obtain more accurate magnetizing relationship, a new method to measure the magnetizing coefficients of the kickers by the real extraction beam was given and the data analysis would also be processed.

## INTRODUCTION

As a high power proton accelerator-based facility [1], the China Spallation Neutron Source (CSNS) consists of an 80 MeV H$^-$ linear accelerator, a 1.6 GeV Rapid Cycling Synchrotron (RCS), a solid tungsten target station, and some instruments for neutron applications [2]. The RCS accelerates the 80 MeV injection beam to the designed energy of 1.6 GeV and extracts the high energy beam to the target. The design goal of beam power for CSNS is 100 kW and can be upgradable to 500 kW [3].

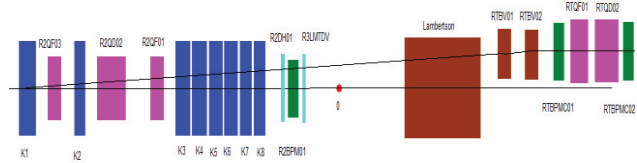

Figure 1: Layout of the RCS extraction system.

The CSNS\RCS has a four-fold lattice with four long straight sections of the injection, extraction, Radio Frequency (RF) and beam collimation [4]. CSNS uses the fast one-turn extraction which is employed in the transfer of beam from RCS to RTBT (the beam transport line from RCS to the target) [5]. Figure 1 shows the layout of the RCS extraction system. It mainly consists of the kicker system and one lambertson magnet. The kicker system includes three tanks and eight kickers: the first tank consists of kicker 1, the second tank consists of kicker 2 and the third tank consists of the last six kickers (kicker 3-8).

___________________________________________
*Work supported by National Natural Science Foundation of China (Project Nos. 11205185)
#huangmy@ihep.ac.cn

In order to obtain the field uniformity and magnetizing relationship of the eight kickers, the magnetic test needs to be done. By some numerical analysis, the magnetic test results can be processed. Then the field uniformity and magnetizing curves can be given and the magnetizing fitting equations can be obtained. However, due to the vacuum errors of the three tanks which consist of eight kickers during the magnetic test, there may be some large measurement errors of the magnetizing relationship. Therefore, during the beam commissioning in the future, new methods to measure more accurate magnetizing relationship of the kickers should be studied and given.

## MAGNETIC TEST OF THE KICKERS

For the extraction system of CSNS, the eight kickers give the extraction beam about 92 mm displacement and 20 mrad deflection in the vertical plane at the extraction point. In order to extract the beam from RCS to RTBT, the eight kickers have different kinds of magnetic field strengths combinations and their magnetizing relationship need to be obtained. Therefore, the magnetic test of the kickers needs to be done.

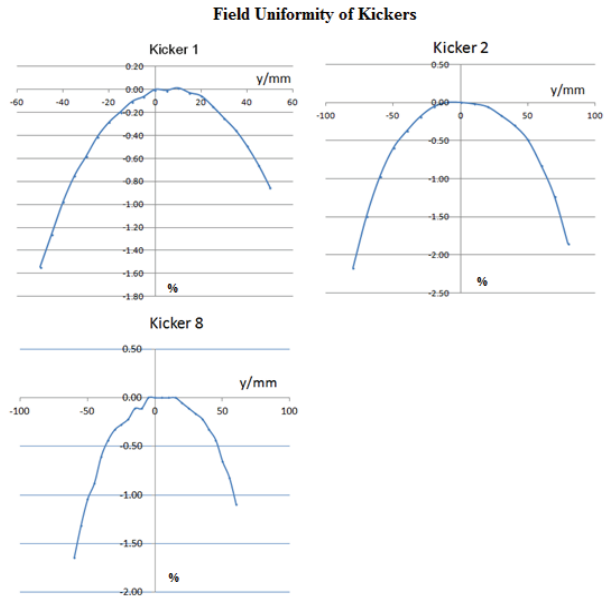

Figure 2: Field uniformity of kicker1, kicker2 and kicker8.

After the magnetic test of the eight kickers, their field uniformity can be obtained, as shown in Fig. 2. It can be known that the field uniformity of different kickers is different. The magnetizing relationship of the kickers also can be obtained after the magnetic test. By using some codes of numerical analysis, the magnetizing curves can be fitted and the magnetizing fitting equations can be given. Table 1 shows the first-order magnetizing fitting equations of the eight kickers.

Table 1: The Magnetizing Fitting Equations of the Eight Kickers

| | Equation (B/T, U/kV) |
|---|---|
| K1 | B = 0.001227×U－2.747×10$^{-4}$ |
| K2 | B = 0.001313×U＋7.956×10$^{-6}$ |
| K3 | B = 0.001266×U－3.269×10$^{-4}$ |
| K4 | B = 0.001265×U－3.228×10$^{-4}$ |
| K5 | B = 0.001330×U＋3.251×10$^{-5}$ |
| K6 | B = 0.001348×U＋3.117×10$^{-5}$ |
| K7 | B = 0.001460×U－3.427×10$^{-4}$ |
| K8 | B = 0.001492×U－6.941×10$^{-4}$ |

## A METHOD TO MEASURE THE MAGNETIZING RELATIONSHIP OF THE KICKERS BY THE REAL EXTRACTION BEAM

Due to the vacuum errors of the three tanks which consist of eight kickers during the magnetic test, there may be some large measurement errors of the magnetizing relationship. Therefore, during the beam commissioning in the future, the magnetizing relationship of the kickers should be re-measured and more accurate magnetizing coefficients should be obtained. In this section, a new method to measure the magnetizing relationship of the kickers by the real extraction beam is studied and the data analysis is also processed.

During the beam commissioning, if the magnetic field strengths combination of the eight kickers is suitable, the beam can be extracted from RCS to RTBT accurately. After the commissioning of RTBV1 and RTBV2, the beam will transport without displacement and deflection after RTBV2. Therefore, the beam positions detected by RTBPMC01 and RTBPMC02 will be almost the same, as shown in Fig. 1.

After the magnetic field strengths combination of the eight kickers and RTBV1 and RTBV2 is suitable, the magnetizing relationship of the kickers can begin to be measured by the real extraction beam. By changing the voltage of kicker "i" ($\Delta U_i$), the beam will transport with displacement and deflection after RTBV2. Then, the beam positions detected by RTBPMC01 and RTBPMC02 will be different. By using the beam displacement deviation and the transfer matrix between RTBPMC01 and RTBPMC02, the beam deflection angle after RTBV2 can be calculated ($\Delta\theta_i$). Therefore, the magnetizing coefficient of kicker "i" can be calculated by

$$k_i = \frac{\Delta B_i}{\Delta U_i} = \frac{\Delta\theta_i \times (B\rho)_{1.6}}{L_{eff,i} \times \Delta U_i}, \qquad (1)$$

where $L_{eff,i}$ is the effective length of kicker "i" and $(B\rho)_{1.6}$ is the magnetic stiffness of the proton beam with 1.6 GeV. By using Eq. (1), the magnetizing coefficient of different kicker can be calculated and the magnetizing relationship will be obtained.

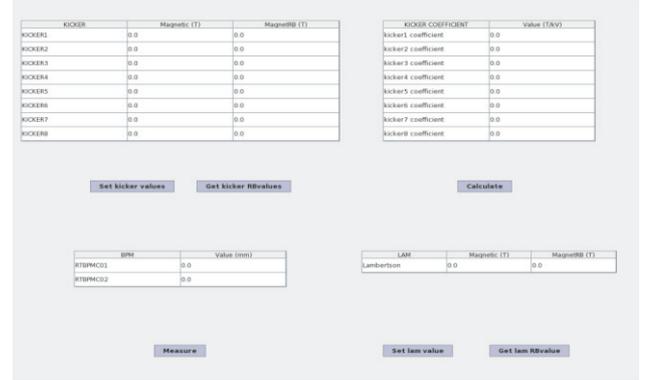

Figure 3: The physics control interface of the extraction system.

Based on the XAL application development environment which was developed initially by SNS laboratory [6], the beam commission software of the extraction system for CSNS was programmed with the Java language [7]. Figure 3 shows the physics control interface of the extraction system. From this figure, firstly, the eight kickers and the lambertson magnet can be controlled and adjusted in the interface. Secondly, the magnetizing coefficients of the eight kickers can be measured and calculated by this software. It can be seen that, after the magnetic field intensities combination of the eight kickers and RTBV1 and RTBV2 is suitable, the beam positions detected by RTBPMC01 and RTBPMC02 will be almost the same. Then, by changing the magnetic field intensity of kicker "i" (its voltage is also adjusted), the beam positions detected by RTBPMC01 and RTBPMC02 will be different. Press the "Calculate" button and the magnetizing coefficient of kicker "i" can be calculated and then given in the table above the "Calculate" button. Therefore, the magnetizing relationship of the kickers can be measured by the real extraction beam during the beam commissioning in the future.

## CONCLUSIONS

In this paper, the magnetic tests of the eight kickers are introduced firstly and the magnetic test results are processed by some numerical analysis. The filed uniformity and magnetizing relationship of the kickers are obtained and then the magnetizing fitting equations are given.

Due to the vacuum errors during the magnetic test, there may be some large measurement errors of the magnetizing relationship of the eight kickers. Therefore, during the beam commissioning in the future, a new method to measure the magnetizing relationship of the kickers by the real extraction beam is studied. In addition, the measurement data is analyzed by the beam commission software of the extraction system and the magnetizing coefficients of the kickers can be calculated.

## ACKNOWLENDGMENTS

The authors want to thank other CSNS colleagues for the discussions and consultations.

## REFERENCES

[1] S. Wang *et al.*, *Chin Phys C*, 33, pp. 1-3, 2009.
[2] CSNS Project Team, *China Spallation Neutron Source Feasibility Research Report*, Chinese Academy of Sciences, 2009 (in Chinese).
[3] J. Wei *et al.*, *Chin Phys C*, 33, pp. 1033-1042, 2009.
[4] M.Y. Huang *et al.*, *Chin Phys C*, 37, p. 067001, 2013.
[5] G.H. Rees, "Extraction", 1993 CAS Accelerator School, CERN 94-04, 1994.
[6] J. Galambos *et al.*, "XAL application programming framework," in *Proc. of ICALEPCS'03*, Gyeongju, Korea, 2003.
[7] M.Y. Huang *et al.*, "Study on the injection beam commissioning software for CSNS/RCS", in *Proc. IPAC2015*, Richmond, VA, USA, pp. 950-952, 2015.